# Temperature dependence of (111) and (110) ceria surface energy

Anastasiia S. Kholtobina[*]

*KTH Royal Institute of Technology, Brinellvagen 23, SE-100 44 Stockholm, Sweden*

Axel Forslund

*Institute for Materials Science, University of Stuttgart, 70569 Stuttgart, Germany*

Andrei V. Ruban

*KTH Royal Institute of Technology, Brinellvagen 23, SE-100 44 Stockholm, Sweden*
*and Materials Center Leoben Forschung GmbH, Roseggerstraße 12, A-8700 Leoben, Austria*

Börje Johansson

*Department of Physics and Astronomy, Division of Materials Theory, Uppsala University, 751 20 Uppsala, Sweden and*
*KTH Royal Institute of Technology, Brinellvagen 23, SE-100 44 Stockholm, Sweden*

Natalia V. Skorodumova

*KTH Royal Institute of Technology, Brinellvagen 23, SE-100 44 Stockholm, Sweden*

[*]anastasiia.kholtobina@gmail.com

High temperature properties of ceria surfaces are important for many applications. Here we report the temperature dependences of surface energy for the (111) and (110) $CeO_2$ obtained in the framework of the extended two-stage up-sampled thermodynamic integration using Langevin dynamics (TU-TILD). The method was used together with machine-learning potentials called moment tensor potentials (MTPs), which were fitted to the results of the ab initio MD calculations for (111) and (110) $CeO_2$ at different temperatures. The parameters of MTPs training and fitting were tested and the optimal algorithm for the ceria systems was proposed. We found that the temperature increases from 0 K to 2100 K led to the decrease of the Helmholtz free energy of (111) $CeO_2$ from 0.78 J/m$^2$ to 0.64 J/m$^2$. The energy of (110) $CeO_2$ dropped from 1.19 J/m$^2$ at 0 K to 0.92 J/m$^2$ at 1800 K. We show that it is important to take anharmonicity into account as simple consideration of volume expansion gives wrong temperature dependences of the surface energies.

## I. INTRODUCTION

Due to its attractive redox, catalytic, electronic, and mechanical properties ceria ($CeO_2$) has become a key material in many modern technologies [1–6]. In particular, the automobile exhaust gas catalysts [1,2], solid oxide fuel cells [5,6] and oxygen storage [7] largely rely on the surface properties of ceria [8]. In these applications ceria works at elevated temperatures [7] and, therefore, the knowledge of the thermodynamic properties of ceria surfaces at high temperatures is of great importance. Such information, however, is still limited, even the data about the energy of different crystallographic surfaces of ceria is scarce. At the same time, the experimental information regarding the properties of bulk ceria is available for different temperatures. For example, ceria heat capacity obtained for the temperature range of 2-900 K using the adiabatic scanning calorimetry and differential scanning calorimetry (DSC) have been reported in a number of studies [9–13]. The high-temperature enthalpy data for bulk ceria in the temperature range of 391-1800 K can also be found in literature [14–16]. Additionally, Hisashige and co-authors have measured the thermal expansion of ceria by thermomechanical analysis (TMA) in the temperature range from 100 to 800 K and the Debye temperature at room temperature by an ultrasonic pulse method [17].

During the past decade a number of ab initio studies devoted to the high-temperature thermodynamic properties of bulk $CeO_2$ were published [7,13,18–20]. The self-consistent ab initio lattice dynamical (SCAILD) method [19], which includes the effects of phonon-phonon interactions using the quasi-harmonic approximation (QHA) provided a description of phonon and thermodynamic properties of bulk ceria at temperatures from 0 K up to 1500-1800 K [21]. Based on the QHA method and phonon calculations the thermodynamic properties, such as heat capacity, isothermal bulk modulus, Gibbs free energy, and coefficient of thermal expansion of $CeO_2$ polymorphs were obtained in the temperature range of 0-1150 K [20]. Morrison et. al. [13] calculated the entropy, enthalpy, and Gibbs functions for bulk ceria at temperatures between 5 K and 400 K using the Perdew, Burke, and Ernzerhof parameterization revised for solids (PBEsol) and the simple Debye model [13]. The obtained Debye temperature ($\Theta_D$), 455 K [13], was in between the previously reported experimental data: 409 K [22], 480 K [17], and reported theoretical values of 481 K [23], 414.5–582.9 K [24]. Additionally, Klarbring et. al. [7] used the temperature dependent effective potential (TDEP) method to investigate several high temperature properties of ceria including thermal expansion [25,26]. Using PBEsol+U, Weck and Kim [23] obtained the crystalline parameters of $CeO_2$ in good agreement with the experimental values. They also calculated the Debye temperatures within the Voigt–Reuss–Hill (VRH) approximation. Niu et. al. [24] obtained the pressure and temperature dependences of the specific heat, Debye temperature, and the thermal expansion coefficient for cubic $CeO_2$ from the Debye–Grüneisen model.

The information about ceria surfaces at elevated temperatures is scarce in experimental publications and totally non-existing in theoretical reports. Zouvelou et. al. [27] described the first experimental determination of the surface energies of polycrystalline $CeO_2$ in the argon atmosphere in the 1473-1773 K temperature range, which were measured to be 1.64-1.47 J/m$^2$. Hayun et. al. [28] determined the surface energy of nanoceria for hydrated and anhydrous samples at room temperature to be 0.81 and 1.16 J/m$^2$, respectively. Hayun et. al. concluded that these surface energies could be attributed to the (111) surface. This assumption was based on the systematic study of the $CeO_2$ nanoparticles of different sizes using high-resolution transmission electron microscopy (HRTEM) [29] and the work by Vyas et. al. [30] showing that $CeO_2$ equilibrium morphologies are dominated by the (111) facets.

Here we report the calculated free energies of the (111) and (110) ceria surfaces at temperatures up to 2100 K. The values were calculated with the extended two-stage upsampled thermodynamic integration using Langevin dynamics (TU-TILD) method [31,32]. This method was previously applied in the free energy calculations of various metallic systems, including vacancy formation free energies [33], stacking fault free energies [31] as well as the surface free energy of TiN [34] and W [35]. In the current study TU-TILD was combined with machine-learning potentials, in particular, the moment tensor potentials (MTP) [36], which were trained on the results of ab initio molecular dynamics (AIMD) calculations [37].

## II. METHODOLOGY

The methodology of the two-stage up-sampled thermodynamic integration using Langevin dynamics was applied in this work to determine the Helmholtz free energies of the ceria bulk and surface supercells and consequently the surface free energies of (111) and (110) ceria

### A. Surface free energy

The slab technique was used to determine the surface free energy $\gamma$ as:

$$\gamma(T) = \frac{F_{slab}(a_T, T) - F_{bulk}(a_T, T)}{2A_T}, \tag{1}$$

where $F_{bulk}(a_T, T)$, and $F_{slab}(a_T, T)$ refer to the Helmholtz free energies of the slab and bulk calculated for the same number of formula units, respectively; $a_T$, is the lattice constant at temperature T and zero pressure, and $A_T$ is the surface area of the slab. Factor 1/2 accounts for the two surfaces of the slab.

The Helmholtz free energies of both bulk and surface supercells (subscripts are omitted in the formula) can be adiabatically decomposed into the following contributions:

$$F(a_T, T) = E(a_T) + F^{vib}(a_T, T), \tag{2}$$

where E denotes the conventional 0 K total energy of the system (either bulk or slab) and $F^{vib}$ the vibrational free energy of the lattice, obtained in the fully anharmonic form using the TU-TILD method.

### B. Anharmonic free energy calculations

We applied the two-stage up-sampled thermodynamic integration using Langevin dynamics method, which treats the interatomic potential as an intermediate reference potential in the thermodynamic integration. The thermodynamic integration is split in two stages: first, from the harmonic to the reference potential and, secondly, from the reference potential to full DFT [38]. In the framework of this study, a modified version of the original TU-TILD method was utilized. The first modification was the usage of an optimized Einstein crystal as the analytic reference to compute the absolute free energy, instead of a quasiharmonic reference. The Einstein crystal is a simple and convenient workaround [39]. The corresponding Einstein frequency can be chosen quite arbitrarily within a reasonable interval specific for this system. Since the Einstein system is used only as an auxiliary reference for thermodynamic integration, the choice does not affect the final result obtained after the integration. The second modification of the original TU-TILD method implemented here was the usage of a machine-learning potential, namely MTP, as an efficient bridge between the analytical reference system and the DFT system.

Following this formalism [40,41], the full vibrational free energy including the anharmonic is obtained as follows:

$$F^{vib} = F^{Einst} + F^{Einst \to MTP} + F^{MTP \to DFT}, \tag{3}$$

where

$$F^{Einst \to MTP} = \int_0^1 d\lambda_1 <E^{MTP} - E^{Einst}>_{\lambda_1}, \qquad (4)$$

$$F^{MTP \to DFT} = \int_0^1 d\lambda_2 <E^{DFT} - E^{MTP}>_{\lambda_2} + <\Delta E>^{UP}. \qquad (5)$$

Where $F^{Einst}$ is the free energy of an optimized Einstein crystal. $E^{Einst}$, $E^{MTP}$, and $E^{DFT}$ are the energies of a particular atomic configuration calculated for the Einstein crystal, calculated with MTP [42] as implemented in the MLIP software [36], and calculated with low-converged DFT parameters, respectively. $<...>_\lambda$ denote the thermodynamic average for particular coupling constant $\lambda$, certain temperature and volume. Finally, $<\Delta E>^{UP}$ is obtained within the free-energy perturbation theory and it accounts for the free energy difference between the low- and well-converged DFT calculations [32].

### C. Machine learning potentials

MTPs are the class of machine-learning (ML) potentials first proposed by Shapeev et. al. for single-component materials [42] and later extended to multi-component systems [43]. MTPs are efficient in combination with the TU-TILD method [1,2]. In the framework of machine learning methods, each considered model should be optimized. In order to avoid overfitting or underfitting during potentials training, the root mean square error (RMSE), calculated between the reference outputs and model predictions, were compared at the end of training process [44].

MTPs represent the energy of an atomic configuration as a sum of the contributions of the local atomic environments of each atom i:

$$E_{tot}^{MTP} = \sum_{i=1}^{n} E_i, \qquad (6)$$

where each contribution $E_i$ is linearly expanded via a set of basis functions,

$$E_i = \sum_\alpha \xi_\alpha B_i^\alpha, \qquad (7)$$

where $\xi = \{\xi_\alpha\}$ are parameters to be found by fitting to the training set.

To train MTPs, an active learning technique can be applied in the framework of the MLIP package [36]. This technique allows one to entrust training set refinement iterations to the computer, thus, completely automating the training set construction. A good training set should include all representative structures, so that the potential does not have to "extrapolate" while searching for stable phases. This goal can be achieved by treating the active learning technique as a generalization of the algorithm proposed for linearly parametrized models by Gubarev [45,46].

The full iteration of the active learning algorithm consists of five steps [47]:

1) training set is constructed from well-converged configurations obtained in DFT-AIMD calculations at the considered temperature;

2) pre-training procedure is implemented for untrained MTP with the defined MTP level and cutoff radius in order to for the first time define current MTP (MTP renewed at each iteration);

3) simulation with the current MTP is performed using the LAMMPS-MLIP interface and the extrapolative configurations are selected; the simulation is stopped when the maximum extrapolation grade [3] is exceeded, and an update of the training set is performed;

4) should new configurations be added to the training set, the total energies of these configurations are calculated with DFT-AIMD and then added to the trained set;

5) MTP is retrained using the updated training set.

The whole procedure should be repeated until no new configurations appear in the third step. If only first two steps are used, we call such a procedure "passive" learning or training since the training set in this case is generated manually and MTP is not "adding" any new configurations [36].

### III. COMPUTATIONAL DETAILS

The DFT calculations were performed using the projector augmented wave (PAW) method [48] as implemented in the Vienna ab initio simulation package (VASP) [49]. The exchange and correlation effects were treated using the Perdew-Burke-Ernzerhof solid $PBE_{sol}$ [50] parametrisation of the generalized gradient approximation (GGA).

Weck et. al. [51] previously demonstrated that $PBE_{sol}$ described the experimental crystalline parameters and properties of $CeO_2$ and $Ce_2O_3$ with good accuracy. For our purpose, the description of cerium oxides within the DFT+U formalism is sufficient [52]. Therefore, the calculations were performed using $PBE_{sol}+U$ energy functional with the Hubbard parameter U of 5 eV applied to the 4f-states of ceria. The PAW potentials with the following electronic configurations were used: Ce 4s4p4f 5d6s and O 2s2p. All calculations were spin polarized with the initial ferromagnetic spin arrangement. The equilibrium lattice parameter of ceria obtained with PBEsol+U was 5.40 Å (0 K), in fair agreement with the experimental value [27].

### A. Bulk free energy calculations

The Helmholtz free energy of the bulk system, $F_{bulk}(a_T, T)$ was obtained using Eq. (2). The convergence parameters were chosen to achieve the accuracy of 1 meV/atom or below. The 0 K total energy, $E_{bulk}(a_T)$, was calculated for the 96-atom supercell built as a $2 \times 2 \times 2$ replication of the 12 atom cubic cell. The total energies were computed for the 12 volumes, equilibrium at different elevated temperatures [7]. For elevated temperatures we used the lattice parameters determined in our previous work [7], which were in good agreement with earlier experimental [25,26] and theoretical data [24]. The plane-wave cutoff was set to 500 eV and the $k$-point mesh was $2 \times 2 \times 2$ [53]. The vibrational free energy $F_{bulk}^{vib}(a_T, T)$ was calculated by the TU-TILD method. The procedure of this calculation consisted of the following three steps:

1) ab initio molecular dynamics runs performed in VASP at 12 temperatures from 450 to 2100 K with 150 K step;
2) 12 moment tensor potentials training in the framework of the active learning algorithm;
3) two-stage up-sampled thermodynamic integration performed using Langevin dynamics with 12 trained MTPs for each considered temperature.

The thermodynamic integration (step 3) includes the following sub-steps corresponding to Eqs.3-5:

*3.1)* calculations of MTP correction to the Einstein crystal model;
*3.2)* calculations of the DFT correction to the MTP free energy;
3.3) additional calculations for the up-sampling term.

All 12 AIMD simulations were run with the Langevin thermostat [54] with the damping parameter of 0.01 fs$^{-1}$. The van-Gunsteren-Berendsen algorithm [55] was used for the integration of the Newton's equations of motion. 1 fs time step was determined to be sufficient for AIMD runs. Each MD run was done for 6000 steps.

For every temperature we trained MTP for it to reproduce the energy and forces of the 6000 atomic configurations for bulk ceria obtained from DFT-AIMD. We did it always in the framework of the active learning approach. The 16$^{th}$ level of MTP and 5 Å cutoff radius were chosen. The resulting root-mean-square error (RMSE) of the energy difference between DFT and MTP was 1.6 meV/atom and RMSE of the force was 0.51 eV Å$^{-2}$ demonstrating a satisfactory reproducibility of the DFT energies and forces by the fitted MTPs.

Having fitted MTPs for the 12 considered temperatures, we could start the two-stage up-sampled thermodynamic integration using Langevin dynamics (Eqs.3-5). The $F^{Einst \rightarrow MTP}$ correction was obtained for the 324-atom supercell built as the $3 \times 3 \times 3$ replication of the 12 atom cubic cell. The convergence of the free-energy correction was below 1 meV/atom for all temperatures. It might be important to use a large enough supercell in this integration in order to capture the contribution of the long wavelength phonons. At every temperature, a dense set of 26 $\lambda_1$ values was used for the integration in Eq. (4). For each $\lambda_1$, LAMMPS MD runs up to 50 000 steps were performed to get statistically well converged results.

The $F^{MTP \rightarrow DFT}$ correction was obtained using the 96-atom supercell. Due to a high computational cost of the calculations at this step only five $\lambda_2$ values (0, 0.25, 0.5, 0.75,1) were chosen for the integration in Eq. (5). For each $\lambda_2$ one MD run with 1000 MD steps was performed. This sampling resulted in the statistical error less than 0.2 meV/atom demonstrating the excellent performance of MTP in reproducing the DFT values. Term $E^{DFT}$ in Eq. (5) was calculated using the plane-wave cutoff of 500 eV and the $2 \times 2 \times 2$ Monkhorst-Pack [53] $k$-point mesh for the 2x2x2 supercell. The calculations of the up-sampling term in Eq. (5), $<\Delta E>^{UP}$, were carried out with the plane-wave cutoff of 700 eV and the $4 \times 4 \times 4$ k-point mesh. It appeared that it was enough to perform 10 up-sampling calculations at each temperature to obtain highly converged energies.

### B. Slab free energy calculations

The Helmholtz free energy of the surface slab, $F_{slab}(a_T, T)$, was obtained using Eq. (2). All the parameters were converged to the accuracy of 1 meV/atom. The 0 K energy $E_{slab}(a_T)$ was calculated for the two (111) surface supercells: $2 \times 2$ and $4 \times 4$ in the *xy* directions, both with the thickness of 9 layers, containing 36- and 144 atoms, respectively. The (110) CeO$_2$ supercell was $2 \times 1$ in the *xy* directions with the thickness of 7 layers (42 atoms). The vacuum was 15 Å thick in all cases. The Monkhorst-Pack [53] $k$-point meshes of $2 \times 2 \times 1$ and $4 \times 4 \times 1$ were used for 4x4 and 2x2 (111) supercells, respectively. For the (110) 2x1 supercell, $4 \times 6 \times 1$ k-point mesh was used. The plane-wave cutoff was 500 eV in all surface calculations.

To calculate $F_{slab}^{vib}(a_T, T)$, MTPs for the surface slabs were fitted in the same manner as bulk-MTPs. The initial DFT AIMD runs were performed for 12 temperatures in the ranges of 450-2100 K for the (111) slab and for 7 temperatures in the ranges of 600-1800 K for the (110) slab. For slab-MTPs, just like for bulk MTPs, the 16th level and cutoff radius of 5 Å were chosen. At each temperature the slab-MTPs were trained to the energies and forces of the 6000 atomic configurations obtained in DFT AIMD runs. The active learning algorithm of MTP training was applied, just like in the case of bulk MTPs. The resulting root-mean-square error (RMSE) of the energy difference between DFT and MTP was 0.3 - 0.5 meV/atom and the RMSE for the force was 0.04 - 0.08 eV Å$^{-2}$.

For the surfaces correction $F^{Einst \rightarrow MTP}$ was calculated using the (111) 4 × 4 and (110) 4 × 8 supercells containing 512 and 672 atoms, respectively. All other parameters were set as for the respective bulk calculations. Correction $F^{MTP \rightarrow DFT}$ for surface slabs was obtained using (111) 4 × 4 and (110) 4 × 2 supercells containing 144 and 168 atoms, respectively. The same set of Lambda values (0, 0.25, 0.5, 0.75, 1) as for the corresponding bulk calculations was used here. At each $\lambda_2$, 1000 step MD runs were carried out with the following DFT parameters: the plane-wave cutoff energy of 500 eV and the 2 × 2 × 1, 4 × 4 × 1 Monkhorst-Pack [53] k-point meshes for the (111) 4 × 4 and (110) 4 × 8 supercells. For up-sampling the following parameters were used: cutoff energy of 700 eV and 6 × 6 × 1 and 4 × 4 × 1 k-point meshes for 36 and 144 atomic (111) $CeO_2$ supercells, respectively, and the 6 × 8 × 1 k-point mesh for 42 atom (110) $CeO_2$ cell. Finally, the surface free energy was calculated according to Eq. (1).

## IV. RESULTS AND DISCUSSION

Here we present the results of the application of the described above methodology to the (111) and (110) ceria surfaces. The temperature dependence of the surface free energy for (111) $CeO_2$ including full anharmonic vibrational contribution is shown in Fig.1. The surface energy decreases from 0.78 J/m² at 0 K to 0.63 J/m² at 2100 K.

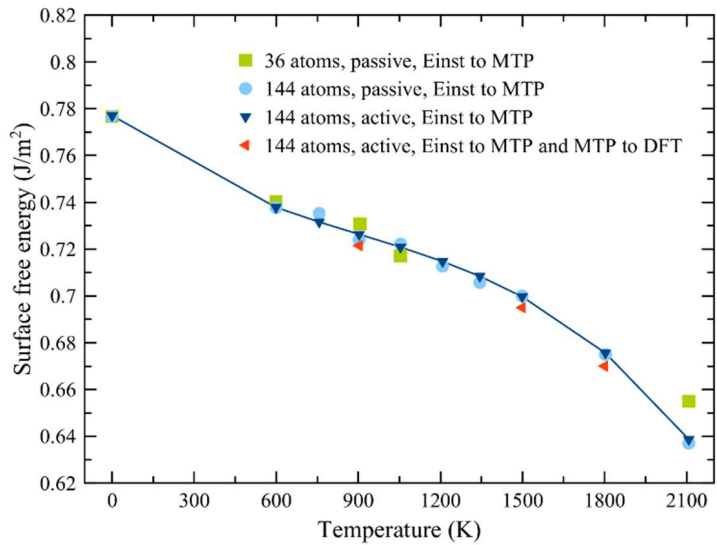

FIG. 1. The surface free energy of (111) ceria surface as the function of temperatures for different supercell sizes, type of MTP training and corrections to the free energy (Einst → MTP and MTP → DFT). Green squares: $\gamma^{Einst} + \gamma^{Einst \rightarrow MTP}$ for the 36 atom unit cell with MTPs obtained from the "passive" learning procedure; blue circles: $\gamma^{Einst} + \gamma^{Einst \rightarrow MTP}$ for the 144 atom unit cell with MTPs trained using "passive" learning; blue triangles: $\gamma^{Einst} + \gamma^{Einst \rightarrow MTP}$ for the 144 atom unit cell with MTPs obtained from "active" learning; the orange triangles: $\gamma_{tot} = \gamma^{Einst} + \gamma^{Einst \rightarrow MTP} + \gamma^{MTP \rightarrow DFT}$ for the 144 atom unit cell using MTPs obtained "active" learning.

In Fig. 1 we compare results obtained for 36 and 144 atoms supercells. It is obvious that already the 36-atom cell allows one to get reasonably accurate results up to high temperatures. Our tests done for the 144-atom cell show that the surface energy does not change whether we use the "active" or "passive" learning procedure for MTP training (Fig.1). Based on this finding, only the last variants of MTPs were used for the 36-atom cell in order to reduce computational cost. Notice that the data in Fig.1 also shows that in this case neither the computationally expensive MTP→DFT correction ( $\gamma^{MTP \rightarrow DFT}$ including the up-sampling procedure) causes any change. This might be explained by a rather high accuracy of our initial MD calculations used to train MTPs in comparison with those used in previous works [34,35] where only the Gamma point and a rather low cutoff energy were used. Therefore, it is not surprising that the authors of Ref [38] obtained a significant MTP→DFT correction providing a noticeable contribution to the final surface free energy.

Based on our tests performed for the (111) surface, we decided to use a 42-atom cell and passive training for MTPs in our calculations for the (110) surface. The MTP→DFT correction, $\gamma^{MTP \to DFT}$ was also neglected. The results for (110) $CeO_2$ are shown in Fig.2 A. Fig.2 B presents the surface energy dependence for (111) $CeO_2$. The (110) surface free energy decreases from 1.19 $J/m^2$ at 0 K to 0.92 $J/m^2$ at 1800 K. Thus, the surface energy decreases in between 0 K and 1800 K is 0.27 $J/m^2$ for (110) $CeO_2$ and 0.15 $J/m^2$ for the close packed (111) surface for the same temperature interval. We notice that it is necessary to take anharmonicity into account in order to adequately describe temperature dependence of ceria surface energy. To demonstrate this point Fig. 2 A, B show the surface energies calculated at 0 K for the volumes corresponding to the considered temperatures.

Fig. 2 A, B also present the results of our transferability tests of the obtained MTPs. The following two types of the transferability tests were performed:

(1) The use of the MTP trained at 1800 K for the evaluation of the surface free energy at lower temperatures. This test was applied for both ceria surfaces. MTPs trained at 1800 K (Fig. 2 A, B) demonstrate good agreement with the results obtained with the MTPs trained at each corresponding temperature (Fig. 2 A, B) both for (111) and (110). The conclusion can be drawn that at least in the case of ceria surfaces it is enough to train MTP at a high temperature and apply it for a range of lower temperatures to estimate the surface free energy with reasonable accuracy.

(2) The use of the individual MTP trained at a particular temperature for (111) ceria for the description of surface free energy of (110) ceria at the same temperature (Fig.2 B) In this case, the best agreement can be found at lower temperatures, for example, 600 K but this approach still can be used for a rough and quick estimation of the (110) surface free energy even at high temperatures.

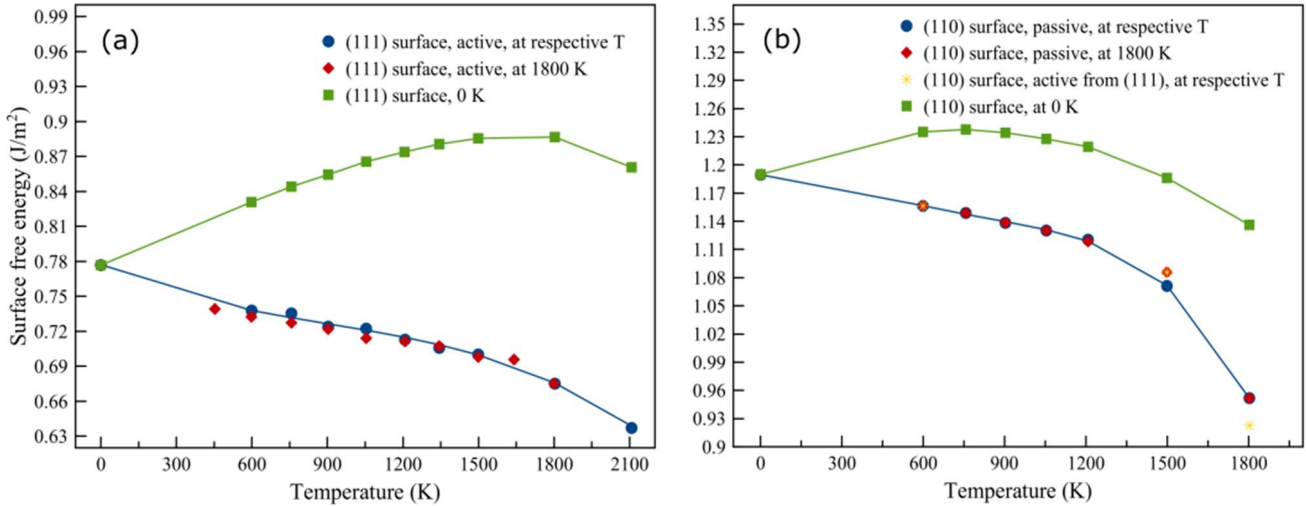

FIG. 2. Surface energies calculated for (111) a) and (110) b) ceria surfaces as a function of the temperature. Green squares: without MTPs at 0 K using lattice parameters respective to finite temperatures.
Blue circles: using the MTPs of corresponding surfaces at respective finite temperatures. Red diamonds: using the 1800 K MTPs of corresponding surfaces. Yellow stars: using the MTPs trained for (111) surface but applied for (110) surface at respective finite temperatures.

Additionally, we tried to use the MTP trained for the (110) surface to describe the (111) surface. This attempt, however, showed no good results even for low temperatures. We note that the MTPs trained for the bulk neither provided any reasonable description of the (111) surface or (110) surface.

Finally, we have shown that the calculation of surface free energy without vibrational contribution to Helmholtz free energy gives us quite different values than the one we obtained taking all contributions into account (green curves in Fig. 2 A, B).

The calculated surface energies can be compared with the available experimental data. Hayun et. al. [28] reported the surface energy value of 0.81 $J/m^2$ at room temperature, which is in fair agreement with our 0K value for the (111) surface (Fig.2 B). Zouvelou et. al. [27] reported the surface energy of polycrystalline $CeO_2$ to be 1.116 – 0.998 $J/m^2$ in the 1473 - 1773 K temperature range, thus, the surface energy decreases in this temperature interval is about 0.12 $J/m^2$ [27]. Note that the type of the surface was not specified in Ref. [27]. Fig.2 A, B demonstrate that for both surfaces the free energy decreases in the temperature interval of 1500 -1800 K. However, in the case of the (111) surface the modest difference of 0.03 J/m$^2$ was obtained whereas for the (110) surface it was 0.12 J/m$^2$, in good agreement with results by Zouvelou et. al. [30].

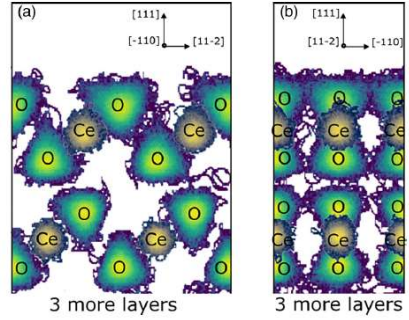

FIG. 3. Atomic trajectories plotted for the (111) surface at 1800 K.  a) projection from the [-110] direction, b) projection from the [11-2] direction. Oxygen atoms are green and cerium atoms are yellow.

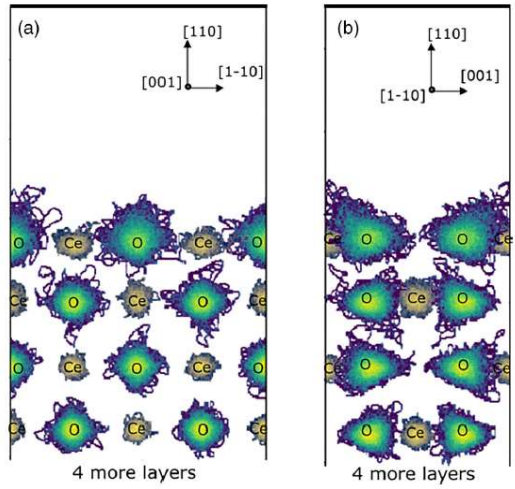

FIG. 4. Atomic trajectories for the (110) ceria surface at 1800 K. a) projected along the in-plane [001], b) projected along the in-plane [110]. Oxygen atoms are green, cerium atoms are yellow.

Figs 3 and 4 present the trajectories of oxygen and ceria atoms in (111) (Fig.3) and (110) surface slabs (Fig.4) obtained from 50000 step MD runs at 1800 K. For both surfaces, as expected, the mobility of surface atoms is larger than that of the bulk atoms. Oxygens shift much further from their crystallographic sites than heavy cerium atoms. In the fluorite structure of ceria, each oxygen is placed in the middle of a tetrahedron of four cerium atoms. The oxygen atom movement is easier through the facet of the cerium tetrahedron than through its edge between two Ce atoms. Under some projection angles Figs. 3 and 4 demonstrate the characteristic triangular shape of the trajectory distributions.

V.   CONCLUSIONS

We demonstrated that the proficient methodology for computing the fully anharmonic surface free energy from ab initio calculations based on the TU-TILD method can be successfully applied to oxide surfaces, in particular to CeO$_2$ (111) and (110) surfaces. The optimal algorithm for the surface free energies calculation in the case of ceria systems has been proposed. It has been shown that for the considered ceria systems active training of the MTPs and utilizing of the increased supercell sizes can be excessive if the original MD calculations are performed with reasonably high accuracy. We have also found that MTP-DFT correction and up-sampling term can be neglected for ceria surfaces again if the initial MD calculations have more k-points than just Gamma point and the cutoff energy of 500 eV.  The surface free energy changing from 0.78 J/m$^2$ at 0 K to 0.64 J/m$^2$ at 2100 K in the case of CeO$_2$ (111) and from 1.19 J/m$^2$ at 0 K to 0.92 J/m$^2$ at 1800 K, in the case of the (110) surfaces. The obtained results are in reasonable agreement with the experimental data by Zouvelou et. al. [27] and Hayun et. al. [28]. It is essential to take anharmonic contributions into account to adequately describe the temperature dependence of the surface free energy of ceria.

**ACKNOWLEDGMENTS**

The computations were enabled by resources provided by the Swedish National Infrastructure for Computing (SNIC) at High Performance Computing Center North and National Computer Center at Linköping University partially funded by the Swedish Research Council (VR-RFI) and through PRACE resources. We acknowledge SNIC and PRACE for awarding us access to Tetralith, Sweden, (Projects SNIC 2021/3-36, SNIC 2022/1-30, and SNIC 2022/22-106) and ARCHER 2, United Kingdom (Project P2021CEO2SB), respectively.

B.J., N.V.S. and A.S.K. acknowledge the financial support from the Carl Trygger Foundation (CTS 20-206). A.F. acknowledges funding from the European Research Council (ERC) under the EUs Horizon 2020 Research and Innovation Program (Grant Agreement No. 865855). A.V.R. also gratefully acknowledges the financial support under the scope of the COMET program within the K2 Center Integrated Computational Material, Process and Product Engineering (IC-MPPE) (Project No. 859480). This program is supported by the Austrian Federal Ministries for Climate Action, Environment, Energy, Mobility, Innovation and Technology (BMK) and for Digital and Economic Affairs (BMDW), represented by the Austrian research funding association (FFG), and the federal states of Styria, Upper Austria, and Tyrol. We are also thankful to Blazej Grabowski for providing software for thermodynamic integration.